\documentclass{article}
\usepackage{spconf,amsmath,graphicx,multirow,array}
\usepackage{epstopdf}
\usepackage{amssymb}
\usepackage{caption}
\usepackage{algorithm} 
\usepackage{algorithmic} 
\usepackage{color}
\usepackage{hyperref}
\usepackage{booktabs}
\usepackage{amsthm,amsmath,amssymb}
\usepackage{cite}
\usepackage{mathrsfs}
\usepackage{subfigure}
\definecolor{note}{RGB}{105, 105, 105} 

\usepackage{setspace}

\title{SICRN: Advancing Speech Enhancement through State Space Model and Inplace Convolution Techniques}
\name{
\begin{tabular}{@{}c@{}}
Changjiang Zhao, Shulin He, Xueliang Zhang\\
\end{tabular}
}

\address{
College of Computer Science, Inner Mongolia University, China\\
\small \texttt{zcj@mail.imu.edu.cn,heshulin@mail.imu.edu.cn,cszxl@imu.edu.cn}
}

\begin{document}
\small
%
\maketitle
\begin{abstract}
Speech enhancement aims to improve speech quality and intelligibility, especially in noisy environments where background noise degrades speech signals.
Currently, deep learning methods achieve great success in speech enhancement, e.g. the representative convolutional recurrent neural network (CRN) and its variants.
However, CRN typically employs consecutive downsampling and upsampling convolution for frequency modeling, which destroys the inherent structure of the signal over frequency.
Additionally, convolutional layers lacks of temporal modelling abilities. 
To address these issues, we propose an innovative module combing a \textbf{S}tate space model and \textbf{I}nplace \textbf{C}onvolution (SIC), and to replace the conventional convolution in CRN, called \textbf{SIC}RN.
Specifically, a dual-path multidimensional State space model captures the global frequencies dependency and long-term temporal dependencies. 
Meanwhile, the 2D-inplace convolution is used to capture the local structure, which abandons the downsampling and upsampling.
Systematic evaluations on the public INTERSPEECH 2020 DNS challenge dataset demonstrate SICRN's efficacy.
Compared to strong baselines, SICRN achieves performance close to state-of-the-art while having advantages in model parameters, computations, and algorithmic delay.
The proposed SICRN shows great promise for improved speech enhancement.
\end{abstract}
\begin{keywords}
Speech enhancement, state space model, deep multidimensional state space model, inplace convolution
\end{keywords}

\vspace{-0.3cm}
\section{Introduction}
\label{sec:intro}
In recent years, the incorporation of deep learning techniques into single-channel speech enhancement methods leads to noteworthy improvements in the speech quality and intelligibility of enhancement systems.
The time domain speech enhancement methods \cite{TCNN,domain_enhancement} utilize neural networks to map noisy speech waveforms and directly enhance speech waveforms.
Frequency domain enhancement techniques \cite{DCCRN} typically employ noise spectral characteristics (e.g., complex spectrum, magnitude spectrum, cepstrum \cite{FFC_SE}, etc.) as inputs for neural models.
The learning target is generally clean speech or a mask (eg, ideal ratio masks\cite{wang2008time}, complex ideal ratio masks \cite{9747888}, etc.). 
Generally speaking, owing to the substantial computational demands of time domain signals and the uncertain representation of feature dimensions, frequency domain methods continue to dominate the landscape of speech enhancement techniques.

In the conventional CRN \cite{CRN} structure, frequency domain enhancement techniques are employed.
The stride of convolutional operation in the frequency dimension is normally set to 2, which shrinks the feature in the frequency dimension.
By stacking the convolutional layers several times, the patterns lying in the frequency dimension are encoded into the channel dimension.
This downsampling operation compromises the inherent feature structure of the original speech, consequently constraining performance.
Therefore, we propose inplace convolution \cite{liu2021inplace,10096918,9747459} method has a significant effect for speech enhancement and acoustic echo cancellation, essentially setting the convolution kernel stride to 1.
Because it does not necessitate the downsampling and upsampling of speech features, It can extract the inherent characteristic information of the original speech.
This extraction occurs without detriment to the amplitude spectrum harmonics and spatial position information.
However, the absence of downsampling operations in inplace convolution makes it challenging to obtain full-band correlations.

Hao et al.\cite{hao2021fullsubnet} propose FullSubNet, a method that achieves effective integration of full-band and sub-band information without downsampling operations, showing remarkable performance.
However, due to the introduction of the full-band model, these methods lead to a large number of overall model parameters, increasing the complexity of the model.
Moreover, while FullSubNet incorporates future frame information and adheres to real-time constraints, it does not strictly adhere to the principles of causal networks.

In this study, we propose  an innovative module that combine multidimensional state space model \cite{nguyen2022s4nd} and inplace convolutions for speech enhancement.
While reducing the amount of parameters and calculations, it solves the problem of speech downsampling destroying the original features and better extracting and fusing local information and global information to improve speech quality and intelligibility.
Specifically, inplace convolution excels in extracting local features and reconstructing speech signals without compromising the integrity of the original feature information. 
However, it lacks full-band correlation.
On the other hand, S4ND proves beneficial for capturing global features while also preserving the integrity of the original feature information.
But S4ND lacks detailed sub-band feature information.
As a result, we propose combining S4ND and inplace convolution to leverage their respective strengths and compensate for each other's weaknesses.
The experimental results demonstrate notably high evaluation scores.
Notably, this achievement is attained with less than 1/2 the parameters and 1/7 the computational complexity of FullSubNet.
Furthermore, SICRN operates without reliance on future frames for enhancement, adhering to the principles of causal networks.
The contributions are as follows:

1. We propose  an innovative module called SIC for speech enhancement by combining S4ND and inplace convolution. 

2. SICRN attains a remarkable level of performance while utilizing merely 2.16 M parameters and 4.24 G/s MACs.


\section{Related Work}
\subsection{S4- State Space Model}
\label{sec:Method Description}

The recently proposed deep neural state-space model(SSM) \cite{gu2021efficiently} advances speech tasks by combining the properties of both CNNs and RNNs. 
The SSM is defined in continuous time using the following equations:
\begin{equation}
   h^{\prime}(t)=A h(t)+B x(t) \label{eq:1}
\end{equation}
\begin{equation}
   y(t)=C h(t)+D x(t)
\end{equation}
To be applied on a discrete input sequence (u0, u1, . . .) instead of continuous function u(t), \eqref{eq:1} must be discretized by a step size $\operatorname{\Delta}$ that represents the resolution of the input.
The discrete SSM is
\begin{equation}
  x_{k}=\overline{\boldsymbol{A}} x_{k-1}+\overline{\boldsymbol{B}} u_{k} \quad y_{k}=\overline{\boldsymbol{C}} x_{k}
\end{equation}
\begin{equation}
  \overline{\boldsymbol{A}}=(\boldsymbol{I}-\Delta / 2 \cdot \boldsymbol{A})^{-1}(\boldsymbol{I}+\Delta / 2 \cdot \boldsymbol{A}) 
\end{equation}
where $\overline{\boldsymbol{A}}$,$\overline{\boldsymbol{B}}$,$\overline{\boldsymbol{C}}$ are the discretized state matrices.According to the conclusion in \cite{gu2021efficiently}, it can be seen that:
\begin{equation}
y_{k}=\overline{\boldsymbol{C A}}^{k} \overline{\boldsymbol{B}} u_{0}+\overline{\boldsymbol{C A}}^{k-1} \overline{\boldsymbol{B}} u_{1}+\cdots+\overline{\boldsymbol{C A} \boldsymbol{B}} u_{k-1}+\overline{\boldsymbol{C} \boldsymbol{B}} u_{k} \label{eq:5}
\end{equation}
\begin{equation}
y=\overline{\boldsymbol{K}} * u
\end{equation}
\begin{equation}
\overline{\boldsymbol{K}}=\left(\overline{\boldsymbol{C} \boldsymbol{B}}, \overline{\boldsymbol{C A B}}, \ldots, \overline{\boldsymbol{C A}}^{L-1} \overline{\boldsymbol{B}}\right)
\end{equation}
In other words, \eqref{eq:5} is a single (non-circular) convolution and can be computed very efficiently with FFTs, provided that $\overline{\boldsymbol{K}}$ is known.
For the specific details of SSM, you can refer to \cite{gu2021efficiently,goel2022s} to understand.

\subsection{S4ND- Multidimensional State Space Model}
The S4 layer was developed for 1-D inputs, which limits its applicability.
In \cite{gu2021efficiently,goel2022s}, the input dimension is 2-D, the shape is (H, T), and the S4 layer is designed as H independent parallel calculations. 
Since there are no correlations in the H dimension, this is limited in speech signal processing.
So S4ND compensates for the correlation in the frequency dimension.
In \cite{nguyen2022s4nd}, the conventional S4 layer was extended to multidimensional signals by turning the standard SSM (1-D ODEs) into multidimensional partial differential equations (PDEs) governed by an independent SSM in each dimension. 

Let $u=u(t^{(1)}, t^{(2)})$ and $y=y(t^{(1)}, t^{(2)})$  be the input and output which are signals $\mathbb{R}^{2} \rightarrow \mathbb{C}$, and $x=(x^{(1)}(t^{(1)}, t^{(2)})$, $x^{(2)}(t^{(1)}, t^{(2)})) \in \mathbb{C}^{N^{(1)} \times N^{(2)}}$ be the SSM state of dimension $N^{(1)} \times N^{(2)}$, where $x^{(\tau)}: \mathbb{R}^{2} \rightarrow \mathbb{C}^{N^{(\tau)}}$.
The 2D SSM is the map u $\mapsto$ y defined by the linear PDE with initial condition x(0, 0) = 0 :
\begin{equation}
\begin{aligned}
\frac{\partial}{\partial t^{(1)}} x(t^{(1)}, t^{(2)}) & =(\boldsymbol{A}^{(1)} x^{(1)}(t^{(1)}, t^{(2)}), x^{(2)}(t^{(1)}, t^{(2)}))+\\ &\phantom{=}\ \   \boldsymbol{B}^{(1)} u(t^{(1)}, t^{(2)}) \\
\frac{\partial}{\partial t^{(2)}} x(t^{(1)}, t^{(2)}) & =(x^{(1)}(t^{(1)}, t^{(2)}), \boldsymbol{A}^{(2)} x^{(2)}(t^{(1)}, t^{(2)}))+\\ &\phantom{=}\ \   \boldsymbol{B}^{(2)} u(t^{(1)}, t^{(2)}) \\
y(t^{(1)}, t^{(2)}) & =\langle\boldsymbol{C}, x(t^{(1)}, t^{(2)})\rangle \label{eq:8}
\end{aligned}
\end{equation}
Note that \eqref{eq:8} differs from the usual notion of multidimensional SSM, which is simply a
map from $u(t) \in \mathbb{C}^{n} \mapsto y(t) \in \mathbb{C}^{m}$ for higher-dimensional n, m $>$ 1 but still with 1 time axis. 
However, \eqref{eq:8} is a map from $u\left(t_{1}, t_{2}\right) \in \mathbb{C}^{1} \mapsto y\left(t_{1}, t_{2}\right) \in \mathbb{C}^{1}$ for scalar input/outputs but over multiple time axes.
When thinking of the input $u(t^{(1)}, t^{(2)})$ as a function over a 2D grid, \eqref{eq:8} can be thought of as a simple linear PDE that just runs a standard 1D SSM over each axis independently
For details, please refer to \cite{nguyen2022s4nd}.
S4ND can be regarded as a convolution kernel with infinite receptive fields in N dimensions.
In the frequency domain signal, the enhanced performance of S4ND-U-Net \cite{ku2023multi} is significant.
Furthermore, both the parameter count and computational load are remarkably low. 
Importantly, S4ND adheres to the principles of causal networks and ensures real-time performance.

\section{METHODOLOGY}
\begin{figure*}[htb] 
\centering

     \label{subfig:mag_impulse}
    \includegraphics[width=14cm]{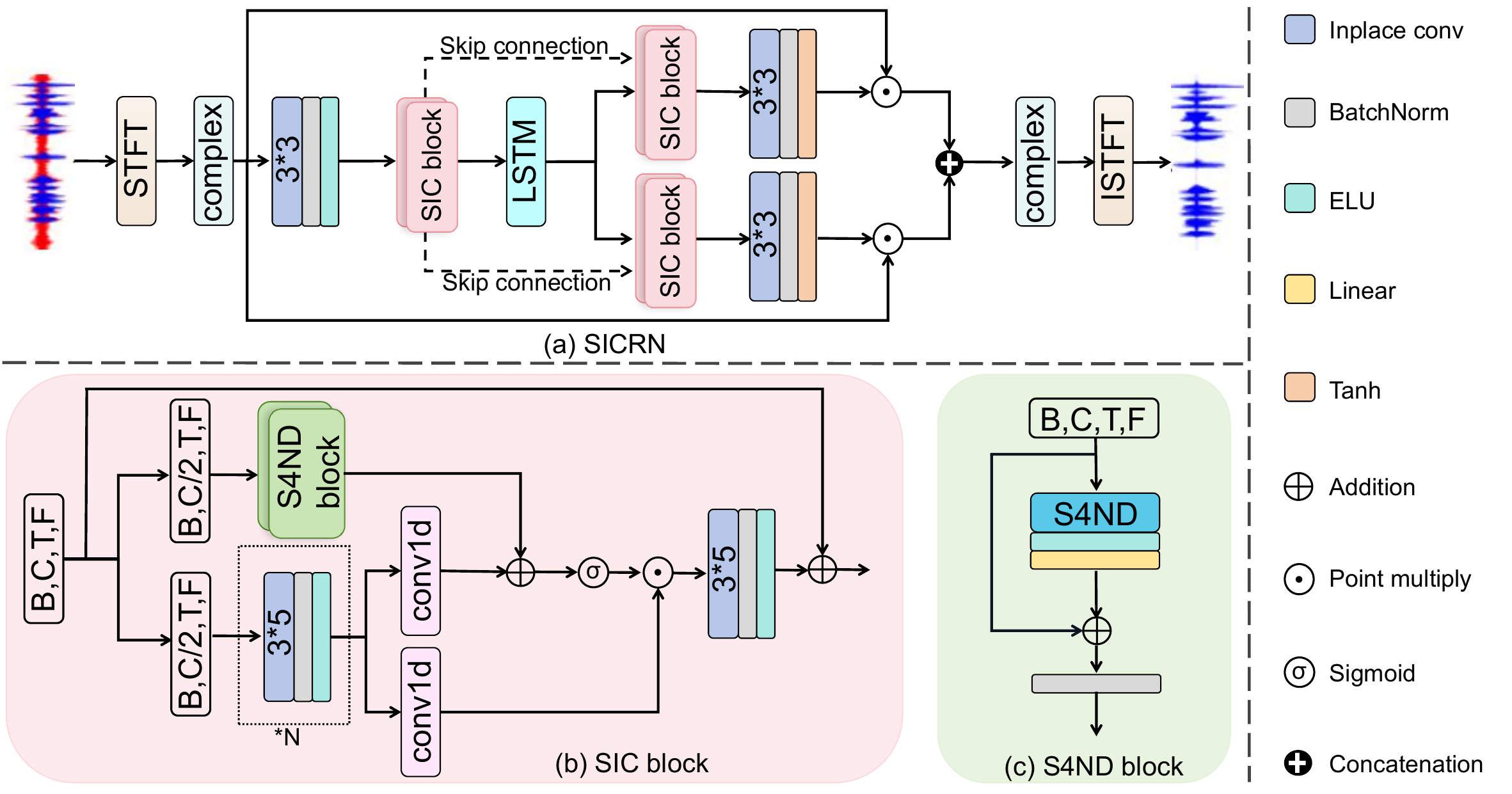}
    
\caption{Overview of the proposed SICRN system}

\label{fig:first}
\end{figure*}
\subsection{SICRN}
SICRN is shown in figure \ref{fig:first}(a).
The comprehensive network architecture adheres to the overall-detailed framework, employing the complex spectrum as its input.
Initially, the feature channels are modified through inplace convolution. Subsequently, the SIC block undertakes preliminary extraction and integration of both global and local features from the real and imaginary components.
Subsequently, the features undergo temporal modeling via a 2-layer LSTM.
In the final step, the SIC block is employed for individual extraction and reconstruction of features from the real and imaginary components. Additionally, it estimates a complex mask, which is then applied through multiplication with the original real and imaginary parts to yield the enhanced feature spectrum.

\subsection{SIC block}
\label{sssec:anchor-voiceprint-features}
The SIC block can function as a convolution kernel, effectively replacing standard convolution kernels, and excels at extracting both local and global features. 
Importantly, the absence of downsampling operations within the entire module ensures the preservation of the original features.

The SIC block is shown in figure \ref{fig:first}(b).
The input channel is bifurcated into two segments. 
The initial 1/2 of the channel employs the inplace convolution kernel to capture local information. 
Following 1D convolution, they serve as the original features and local attention features, respectively.
The latter 1/2 of the channel feeds into the S4ND layer for global information extraction, which is then harmonized with local attention features.
The attention map is derived via the sigmoid activation function and subsequently multiplied with the  original feature, thereby facilitating the extraction and fusion of local and global features.
The specific calculation process is as follows:
\begin{equation}
 \begin{split}
    X_{0 \sim \frac{c}{2}}^{L}=C_{1d}(\operatorname{IC}(X_{0 \sim \frac{c}{2}})
\end{split}
\end{equation}
\begin{equation}
 \begin{split}
    X_{\frac{c}{2} \sim c}^{R}=\operatorname{S}(X_{\frac{c}{2} \sim c})
\end{split}
\end{equation}
\begin{equation}
 \begin{split}
     A T m a p=\sigma(C_{1d}(\operatorname{IC}(X_{0 \sim \frac{c}{2}}))+X_{\frac{c}{2} \sim c}^{R})
\end{split}
\end{equation}
\begin{equation}
 \begin{split}
    \mathrm{X}=X_{0 \sim \frac{c}{2}}^{L} \cdot A T m a p
\end{split}
\end{equation}
Where, $X_{0 \sim \frac{c}{2}}^{L}$ denotes the local features, while $X_{\frac{c}{2} \sim c}^{R}$ represents the global features. 
$X_{0 \sim \frac{c}{2}}$ corresponds to the features of the first half of the channels, and $X_{\frac{c}{2} \sim c}$ pertains to the features of the latter half of the channels. 
$\operatorname{IC}(\cdot)$ denotes the inplace convolution operation, and $\operatorname{S}(\cdot)$ signifies the convolution kernel used in the S4ND layer.
The symbol $\sigma$ represents the sigmoid activation function.
'$A T m a p$' denotes the attention map and $C_{1d}(\cdot)$ denotes the conv1d.

\subsection{S4ND block}
Figure \ref{fig:first}(c) shows the S4ND block.
Initially, the S4ND is employed to extract global features. This extraction process is followed by passage through an ELU activation function, and output via a linear layer.
Subsequently, a residual connection is implemented to address potential problem related to gradient vanishing or exploding.
Finally, a batch normalization layer is applied to generate the final output.
The rationale behind choosing S4ND for global feature modeling is as follows:

1. The global modeling capacity of S4ND surpasses that of LSTM, while maintaining a smaller parameter count and computational load.

2. Given that S4 processes elements independently within the frequency dimension, it isn't ideally suited for processing frequency domain information. 
This limitation is addressed by the utilization of S4ND.

Furthermore, as demonstrated in \cite{nguyen2022s4nd}, experimental comparisons between S4ND and 2D convolutions reveal that S4ND outperforms the latter. Consequently, we opt for S4ND as the method to extract global feature information.
\subsection{Loss function}
\label{sssec:Loss-function}
We apply a scale-invariant signal-to-noise ratio (SI-SNR) \cite{luo2019conv} loss, which is a time domain loss function as follows:
\begin{equation}
\mathbf{s}_{\text {target }}=\frac{\langle\hat{\mathbf{s}}, \mathbf{s}\rangle \mathbf{s}}{\|\mathbf{s}\|^2} 
\end{equation}
\begin{equation}
\mathbf{e}_{\text {noise }}=\hat{\mathbf{s}}-\mathbf{s}_{\text {target }}
\end{equation}
\begin{equation}
\mathcal{L}_{\text {si-snr }}=10 \log _{10} \frac{\left\|\mathbf{s}_{\text {target }}\right\|^2}{\left\|\mathbf{e}_{\text {noise }}\right\|^2}
\end{equation}
where $\hat{\mathbf{s}} \in \mathbb{R}^{1 \times T}$ and $\mathbf{s} \in \mathbb{R}^{1 \times T}$ refer to the estimated and clean sources, respectively, and $\|\mathbf{s}\|^2=\langle\mathbf{s}, \mathbf{s}\rangle$ denotes the signal power.

\section{EXPERIMENTAL Setup}
\label{sec:EXPERIMENT}

\subsection{Datasets}
\label{ssec:ataset}
We evaluated the SICRN on the DNS Challenge (INTERSPEECH 2020) dataset \cite{reddy2020interspeech}. 
The clean speech set includes over 500 hours of clips from 2150 speakers.
The noise dataset includes over 180 hours of clips from 150 classes. 
To make full use of the dataset, we simulate the speech-noise mixture with dynamic mixing during model training. 
In detail, before the start of each training epoch, 75\% of the clean speeches are mixed with randomly selected room impulse responses (RIR) from (1) the Multichannel Impulse Response Database \cite{hadad2014multichannel} with three reverberation times (T60) 0.16s, 0.36s, and 0.61 s. (2) the Reverb Challenge dataset \cite{kinoshita2016summary} with three reverberation times 0.3 s, 0.6 s and 0.7 s. 
After that, the speech-noise mixtures are dynamically generated by mixing the clean speech (75\% of them are reverberant) and noise with a random SNR in between -5 and 20 dB. 
The DNS Challenge provides a publicly available test dataset, including two categories of synthetic clips, i.e., without and with reverberations. 
Each category has 150 noisy clips with SNR levels distributed in between 0 dB to 20 dB. We use this test dataset for evaluation.

\subsection{Configuration}
\begin{table*}[h]
\small %
\footnotesize
\centering
\captionsetup{font=small}
\caption{The performance in terms of WB-PESQ [MOS], NB-PESQ [MOS], STOI [\%], and SI-SDR [dB] on the DNS challenge test dataset.}
\begin{tabular}{ccccccccccccc}
\bottomrule
\multirow{2}{*}{Method} & \multirow{2}{*}{\begin{tabular}[c]{@{}c@{}}\#Para\\ (M)\end{tabular}} & \multirow{2}{*}{\begin{tabular}[c]{@{}c@{}}Look Ahead\\ (ms)\end{tabular}} & \multicolumn{4}{c}{With Reverb}    &  & \multicolumn{4}{c}{Without Reverb} \\ \cline{4-7} \cline{9-12} 
                        &                                                                       &                                                                            & WB-PESQ & NB-PESQ & STOI  & SI-SDR &  & WB-PESQ & NB-PESQ & STOI  & SI-SDR \\ \hline
Noisy                   & -                                                                     & -                                                                          & 1.822   & 2.753   & 86.62 & 9.033  &  & 1.582   & 2.454   & 91.52 & 9.071  \\
NSNet {\cite{xia2020weighted}}             & 5.1                                                                   & 0                                                                          & 2.365   & 3.076   & 90.43 & 14.721 &  & 2.145   & 2.873   & 94.47 & 15.613 \\
DTLN {\cite{westhausen2020dual}}              & 1.0                                                                   & -                                                                           &-         & 2.700    & 84.68 & 10.530  &  & -        & 3.040    & 94.76 & 16.340  \\
Conv-TasNet {\cite{koyama2020exploring}}       & 5.08                                                                  & 33                                                                         & 2.750   &-         &-       & -       &  & 2.730    & -        & -      &-        \\
DCCRN-E {\cite{hu2020dccrn}}           & 3.7                                                                   & 37.5                                                                       & -        & 3.077   &-       &-        &  &-         & 3.266   &-       &-        \\
PoCoNet {\cite{isik2020poconet}}           & 50                                                                    & -                                                                           & 2.832   & -        &-       & -       &  & 2.748   &-         & -      & -       \\
FullSubNet {\cite{hao2021fullsubnet}}       & 5.64                                                                  & 32                                                                         & 2.969   & 3.473   & 92.62 & 15.750  &  & 2.777   &3.305  & 96.11 & 17.290  \\ \hline 
SICRN            & 2.16                                                                  & 0                                                                          & 2.891        &3.433         &92.59       &15.137        &  & 2.624        &3.233         &95.83       &15.998       \\
\bottomrule
\end{tabular}
\label{tab:speech1}
\end{table*}

For STFT ,we adopt 510/160 for win-length/hop-length,the analysis is Hanning window.
We use 510-point discrete Fourier transform (DFT) to extract 256-dimensional complex spectra for 16 kHz sampling rate.
The model is optimized by Adam.
The initial learning rate is 0.0002 and halved when the validation loss of four consecutive epochs no longer decreased. 
When training the model, the specific settings are as follows: there are 2 layers of SIC blocks, with channel sizes of 16 and 32, respectively, for each layer. 
Additionally, there are 2 LSTM layers in the middle. Within the SIC block, there are 3 inplace convolution layers, and 4 layers of S4ND blocks.
For the integrity of the experiment, the 4-layer S4ND blocks in SIC block are replaced with 4-layer inplace convolution, which is named IICRN.

In order to verify the effectiveness of the proposed method, we select FullSubNet as the baseline models.
At the same time, compare according to the same evaluation standard of FullSubNet.
In addition, we also compared with the topranked methods in the DNS challenge (INTERSPEECH 2020), including NSNet \cite{xia2020weighted}, DTLN \cite{westhausen2020dual}, Conv-TasNet \cite{koyama2020exploring}, DCCRN \cite{hu2020dccrn} and PoCoNet \cite{isik2020poconet}.

\section{EXPERIMENT RESULTS and Analysis}
In this section, We compare the proposed SICRN network with other baseline systems across various metrics, including SI-SDR, STOI, NB-PESQ, WB-PESQ, “\#Para” and “Look Ahead” , and show the results on the DNS challege test dataset.
 “\#Para” and “Look Ahead” in the table respectively represent the parameter amount of the model and the length of used future information.
 “With Reverb” means that the noisy speeches in the test dataset have not only noise but also a certain degree of reverberation. 
 “Without Reverb” means that the noisy speeches in the test dataset have only noise.

Table \ref{tab:speech1} provides a comprehensive evaluation of several methods. In the “With Reverb” column, SICRN outperforms the majority of models, ranking second only to FullSubNet, with only a negligible 0.03\% difference in the STOI score.
In the “Without Reverb” column, SICRN's evaluation score also stands impressively high, with only a slight decrease compared to FullSubNet.
Furthermore, as demonstrated in “\#Para” and “Look Ahead”, SICRN achieves this high level of performance with a relatively small number of parameters and without relying on future frames for enhancement.
\begin{table}[htbp]
\small %
  \footnotesize
  \centering
  \caption{Comprehensive comparison with FullSubNet.}
    \begin{tabular}{cccc}
    \toprule
    Method & \#Para(M) & MACs(G/s)  & Look Ahead(ms)  \\
    \midrule
    FullSubNet & 5.64  & 30.84  & 32  \\
    SICRN & \textbf{2.16}  & \textbf{4.24}  & \textbf{0} \\
    \bottomrule
    \end{tabular}%
  \label{tab:speech4}%
\end{table}%

Table \ref{tab:speech4} provides a detailed comparison of parameter count, computational complexity, and Look Ahead between SICRN and FullSubNet, highlighting the best scores of each case in bold.
It's evident that SICRN utilizes only 2.16 M parameters and 4.24 G/s MACs. 
In stark contrast, FullSubNet employs 5.64 M parameters and exhibits a computational complexity of 30.84 G/s. 
Therefore, SICRN stands out for its significantly lower parameter count and computational complexity while delivering remarkable performance.
Most notably, SICRN leverages solely the current frame and the past frame for enhancing the current frame, a strategy that presents distinct advantages and untapped potential when compared to FullSubNet.

\begin{table}[htbp]
\small
  \footnotesize
  \centering
  \caption{Ablation experiment(With Reverb).}
    \begin{tabular}{cccccc}
    \toprule
    Method  & WB-PESQ & NB-PESQ  & STOI & SI-SDR  \\
    \midrule
    mixture   & 1.822  & 2.753  & 86.62  & 9.033  \\
    \midrule
    IICRN   & 2.797  & 3.378  & 91.71  & 14.929   \\
    SICRN  & \textbf{2.891}  & \textbf{3.433}  & \textbf{92.59}  & \textbf{15.137}\\
    \bottomrule
    \end{tabular}%
  \label{tab:speech2}%
\end{table}%

\begin{table}[htbp]
  \small %
  \footnotesize
  \centering
  \caption{Ablation experiment(Without Reverb).}
    \begin{tabular}{cccccc}
    \toprule
    Method  &WB-PESQ &NB-PESQ  &STOI &SI-SDR  \\
    \midrule
    Noisy  & 1.582  & 2.454  & 91.52  & 9.071  \\
    \midrule
    IICRN   & 2.596  & 3.218  & 95.56  & 15.795   \\
    SICRN   & \textbf{2.624}  & \textbf{3.233}  & \textbf{95.83}  & \textbf{15.998}   \\
    \bottomrule
    \end{tabular}%
  \label{tab:speech3}%
\end{table}%

To elucidate the role of S4ND, we conducted an ablation experiment to substantiate the benefits arising from the synergy between S4ND and inplace convolution.
Displayed in Table \ref{tab:speech2} and Table \ref{tab:speech3}, they represent the test outcomes for the datasets with reverberation and without reverberation, respectively.
The experiments demonstrate a substantial performance enhancement upon the integration of S4ND, affirming the efficacy of combining S4ND and inplace convolution for feature extraction.
Notably, the enhancement effect is particularly pronounced in the test set with reverberation. 
While the performance of IICRN is slightly inferior to SICRN, it still demonstrates a commendable level of effectiveness. This observation further underscores the advantages of employing inplace convolution to avoid downsampling speech features.

\section{CONCLUSIONS}
\label{sec:CONCLUSIONS}
we propose an innovative model combing a state space model and inplace convolution, called SICRN.
This network avoids downsampling operations throughout its architecture and combines multidimensional state space model and inplace convolution techniques to extract and integrate both global and local features.
The experimental results demonstrate superior performance achieved with fewer parameters and computational resources, all without the need for future frame information.

\textbf{Acknowledgments}: This research was partly supported by the China National Nature Science Foundation (No. 61876214).


\newpage
\bibliographystyle{IEEEbib}
\bibliography{strings,refs}

\end{document}